# Customizable Avatars with Dynamic Facial Action Coded Expressions (CADyFACE) for Improved User Engagement

Megan A. Witherow, Crystal Butler, Winston J. Shields, Furkan Ilgin, Norou Diawara, Janice Keener, John W. Harrington, and Khan M. Iftekharuddin

***Abstract*—Customizable 3D avatar-based facial expression stimuli may improve user engagement in behavioral biomarker discovery and therapeutic intervention for autism, Alzheimer's disease, facial palsy, and more. However, there is a lack of customizable avatar-based stimuli with Facial Action Coding System (FACS) action unit (AU) labels. Therefore, this study focuses on (1) FACS-labeled, customizable avatar-based expression stimuli for maintaining subjects' engagement, (2) learning-based measurements that quantify subjects' facial responses to such stimuli, and (3) validation of constructs represented by stimulus-measurement pairs. We propose Customizable Avatars with Dynamic Facial Action Coded Expressions (CADyFACE) labeled with AUs by a certified FACS expert. To measure subjects' AUs in response to CADyFACE, we propose a novel Beta-guided Correlation and Multi-task Expression learning neural network (BeCoME-Net) for multi-label AU detection. The beta-guided correlation loss encourages feature correlation with AUs while discouraging correlation with subject identities for improved generalization. We train BeCoME-Net for unilateral and bilateral AU detection and compare with state-of-the-art approaches. To assess construct validity of CADyFACE and BeCoME-Net, twenty healthy adult volunteers complete expression recognition and mimicry tasks in an online feasibility study while webcam-based eye-tracking and video are collected. We test validity of multiple constructs, including face preference during recognition and AUs during mimicry.**

This work is supported in part by the National Science Foundation Graduate Research Fellowship under Grant Nos. 1753793 and 2139907, and in part by the Research Computing clusters at Old Dominion University under National Science Foundation Grant No. 1828593.
*(Corresponding author: M. A. Witherow)*

This work involved human subjects in its research. Approval of all ethical and experimental procedures and protocols was granted by the Old Dominion University Institutional Review Board under Application No. 1424272 and the Eastern Virginia Medical School Institutional Review Board under Application No. 19-06-EX-0152.

M. A. Witherow, W. J. Shields, F. Ilgin, and K. M. Iftekharuddin are with the Vision Lab, Department of Electrical & Computer Engineering, Old Dominion University, Norfolk, VA 23529 USA (e-mail: mwith010@odu.edu; winstonjshields@gmail.com; furilgin@gmail.com; kiftekha@odu.edu).

C. Butler is an independent certified FACS Coder, Ellicott City, MD 21043 USA (e-mail: crystal@crystal-butler.com).

N. Diawara is with the Department of Mathematics & Statistics, Old Dominion University, Norfolk, VA 23529 USA (e-mail: ndiawara@odu.edu).

J. W. Harrington and J. Keener are with the Department of Pediatrics, Eastern Virginia Medical School and Children's Hospital of The King's Daughters, Norfolk, VA 23507 USA (e-mail: john.harrington@chkd.org; janice.keener@chkd.org).

## I. INTRODUCTION

OVER the past twenty years and now entering the age of the metaverse, 3D avatars have become increasingly prominent and effective tools for health applications. In clinical settings, avatars have been broadly applied for neuro- and motor-rehabilitation [1] of patients, such as those who have suffered from stroke [2], cerebral palsy [3], brain injury [4], Parkinson's disease [5], Alzheimer's disease and dementia [6], [7]. Furthermore, social avatars have aided discovery of potential behavioral biomarkers and development of therapeutic interventions for social anxiety disorder [8], depression [9], [10], schizophrenia [11], and autism spectrum disorder (ASD) [12], [13], [14], [15]. Given the ubiquity of facial expressions in daily life and their relevance to psychosocial health (e.g., facial palsy [16], depression [9], [10], social anxiety [8], ASD [12], [13], [14]), facial expressions have become important targets for avatar-based health applications. Thus, important design considerations have been identified to ensure that 3D avatars are valid and engaging [17], [18].

Facial expressions of 3D avatars are often used as stimuli in studies of intervention efficacy or behavioral biomarker discovery [9], [10], [13], [14], [15]. Such studies incorporate tasks to elicit and measure constructs related to facial expressions. The typical setting involves eliciting a response using avatar-based stimuli, capturing the response with a sensor, and extracting relevant measurements from the raw sensor data. To capture perception and production of facial expressions, the applicable sensors are eye-trackers and video cameras, respectively [9], [10], [13], [14], [15]. Measures such as the percentage duration of gaze fixations to areas of interest (AOIs) within the stimuli have been used to study perception [19]. To assess production, the Facial Action Coding System (FACS) [20] provides a taxonomy of action units (AUs) that describe individual constituent movements of the face. Machine and deep learning approaches may be used to detect AUs from facial images [21], [22], [23], [24], [25], [26], [27], [28]. Finally, evaluating the construct validity of these tasks, i.e., whether the intended construct is elicited and measured, is an important precursor for well-designed studies of intervention efficacy or behavioral biomarker discovery [19], [29].

In the sections to follow, we review related work on important design considerations for 3D avatar-based facial expression stimuli, automatic detection of FACS AUs, and construct validity. Then, we propose (1) dynamic, FACS-labeled stimuli for perception and production of facial expressions, rendered on customizable 3D avatars, (2) a new deep learning-based AU detector for measurement of subjects' facial responses, and (3) construct validity of proposed stimuli and measurements based on two tasks (recognition and mimicry) completed by 20 healthy adult volunteers.

### A. Related Work

*1) Design Considerations for 3D Avatar-based Facial Expression Stimuli:* Securing and maintaining user engagement is a key challenge for avatar-based health applications [17]. Recently, avatar customization has been identified as an effective means of improving engagement [17]. Avatar customization has been shown to increase engagement and enjoyment in social [30], [31], procedural [32], creative [32], and cognitive tasks [17], including interventions for physical [33], [34] and mental health [17], [31]. Avatar realism is another important factor influencing engagement. Hyper-realistic avatars may trigger the uncanny valley effect, a phenomenon where objects with increasingly realistic human appearances evoke uneasiness or revulsion, causing users to disengage [18]. Furthermore, several studies find that users prefer to interact with semi-realistic avatars [18], [35]. Avatars may embody a humanoid form to varying degrees from 'talking heads' to full body representations. Full body representations have been shown to improve dyadic interactions with avatars [36]. Facial expression stimuli may be rendered statically as still images or dynamically as animations from neutral to peak expression. While some studies [12], [37] use static facial expressions due to accessibility of widely used, validated stimuli sets [38], [39], it has been established from both neuroimaging and behavioral perspectives that dynamic expressions are more salient than static expressions, and show increased activity in face processing regions of the brain [40]. Thus, dynamic facial expressions play a pivotal role in assessing relevant differences between control individuals and individuals with a diagnosis in biomarker discovery studies (e.g., depression [41], Moebius syndrome [42], ASD [40]).

To study the effect of 3D avatar-based stimuli on perception or production of facial expressions, it is important to ensure that the avatar accurately renders the target expressions by having the expressions evaluated and labeled with AUs by FACS experts. This labeling may be especially critical for studies of expression production, where constructs may be defined based on a one-to-one correspondence between the avatar's AUs and the participant's AUs. While several methods for transferring AUs to arbitrary avatar faces have emerged, e.g., [43], [44], these methods are not guaranteed to accurately reproduce the target AUs. Therefore, avatar models and avatar-generation platforms that have been evaluated by FACS experts, such as MiFace [45], HapFACS [46], FACSGen [47], FACSHuman [48], and García et al.'s avatars [49], are preferred. While all of these existing avatars and avatar-generation platforms support dynamic animations, they are limited in that they either lack customization capabilities [45], [49], rely on commercial software [46], [47], and/or are rendered as a disembodied floating head or face [45], [47], [48], which may break immersion. Additionally, García et al.'s avatars [49] are hyper-realistic, which may trigger the uncanny valley effect [18]. Given these limitations, there still exists a need for customizable, dynamic 3D avatars and avatar-based facial expression stimuli that have been evaluated and labeled with AUs by FACS experts.

*2) Automatic Detection of FACS AUs:* While eye-tracking based measures of facial expression perception require only straightforward mathematical operations [19], automatic AU detection from images is more challenging. Facial expressions consist of multiple AUs occurring simultaneously in various localized areas of the face. Thus, AU detection is a multi-label problem, where each facial image is assigned one or more AUs. AU detection methods either train individual binary classifiers to detect the presence or absence of each AU or train a single model to detect multiple AUs at once [23], [24], [25], [26], [28]. The latter approach, referred to as multi-label learning, is considered superior due to its computational efficiency and ability to take relationships between AUs into account [23], [24], [25], [26], [28]. In addition to modeling the relationships between AUs, state-of-the-art multi-label learning approaches often incorporate methods for focusing on relevant areas of the input or features using saliency maps [28], attention [27], or patch/region learning [21], [26]. Multi-label learning approaches may also benefit from multi-task learning of other tasks related to the face (e.g., landmark prediction [27], facial expression classification [50], valence-arousal estimation [50]) and from feature fusion (e.g., saliency maps [28], geometric features [51]).

A drawback of state-of-the-art multi-label AU detection approaches is that they do not independently predict left and right activations of bilaterally located AUs, which may be useful for health applications. For example, Dell'Olio et al. [52] recently propose FaraPy, an augmented reality mirror therapy for patients with facial paralysis. Asymmetrical AU activation is characteristic of facial palsy or paralysis, e.g., due to stroke, Parkinson's, Bell's Palsy, etc. [52], and has also been observed among individuals diagnosed with ASD [53], [54], [55].

Recently, Bar et al. [56] present an approach based on convex geometry for identifying significant correlations among a large number of features using a mixture of beta distributions (betaMix). The betaMix approach relies upon Theorem 1.1 from [57], which shows that the sine squared of the angles between randomly drawn features in high dimensional space follows a beta distribution. The betaMix approach [56] has been applied to facial expression classification in [58] for decomposing geometric landmark-based features (e.g., distances between pairs of landmarks) into sets associated with expressions, identity, and age groups (children and adults). In [58], learning takes place in two separate steps. First, betaMix [56] is fit using the EM algorithm to learn correlations between already extracted landmark-based features and three factors (expressions, identity, and age groups). The resulting graph is used to select expression-correlated features that are invariant to age and identity. Then, in the second step, the betaMix-selected features are fused with deep learning-based features to fit the expression classifier. Thus, feature extraction, selection, and expression learning steps are not trained end-to-end. Intermediate steps lack supervision from the class labels and thus may not be fully optimized for the expression classification task. Given the success with facial expression classification, we hypothesize that the aforementioned theorem [57] may be adapted into a loss function for simultaneous, end-to-end learning of correlations among AUs and features, while discouraging dependence on identity, which is not addressed by present multi-label AU detection approaches.

*3) Construct Validity:* Construct validity may be determined by assessing whether the expected response is elicited in a healthy control group. For example, this approach has been used to evaluate constructs related to candidate eye-tracking biomarkers for ASD [19]. Since neurotypical individuals are known to prefer to attend to faces when viewing social stimuli, Shic et al. [19] test for face preference (percentage of gaze duration to the face AOI vs. random gaze) among neurotypical controls to determine construct validity.

### *B. Contributions*

To address the limitations of currently available 3D avatar-based facial expression stimuli, we propose Customizable Avatars with Dynamic Facial Action Coded Expressions (CADyFACE) for user engagement. To detect AUs elicited by CADyFACE, we propose a deep neural network for novel Beta-guided Correlation and Multi-task Expression learning (BeCoME-Net). We further conduct a feasibility study to evaluate the construct validity of CADyFACE and BeCoME-Net AU measurements. Our contributions are as follows:

- CADyFACE incorporates six avatar models representing different genders and races with customizable hair color, eye color, skin tone, and clothing. For each CADyFACE model, six facial expressions (anger, disgust, fear, happy, sad, and surprise) have been posed and labeled by a certified FACS expert with over 600 hours of coding experience.
- We propose a novel beta-guided correlation loss for BeCoME-Net that encourages features to be correlated with AUs while discouraging correlation with subject identity. For richer representation learning, BeCoME-Net fuses geometric landmark and deep learning-based texture features while jointly learning AU detection and expression classification tasks. We consider variants of BeCoME-Net for bilateral and unilateral AU detection. We compare BeCoME-Net with state-of-the-art AU detection methods on two benchmark data sets.
- We conduct an online feasibility study of 20 healthy adult participants to evaluate the construct validity of the proposed CADyFACE stimuli and BeCoME-Net AU measurements. Participants complete two facial expression-related tasks, recognition and mimicry, while facial video and webcam-based eye-tracking data are collected.

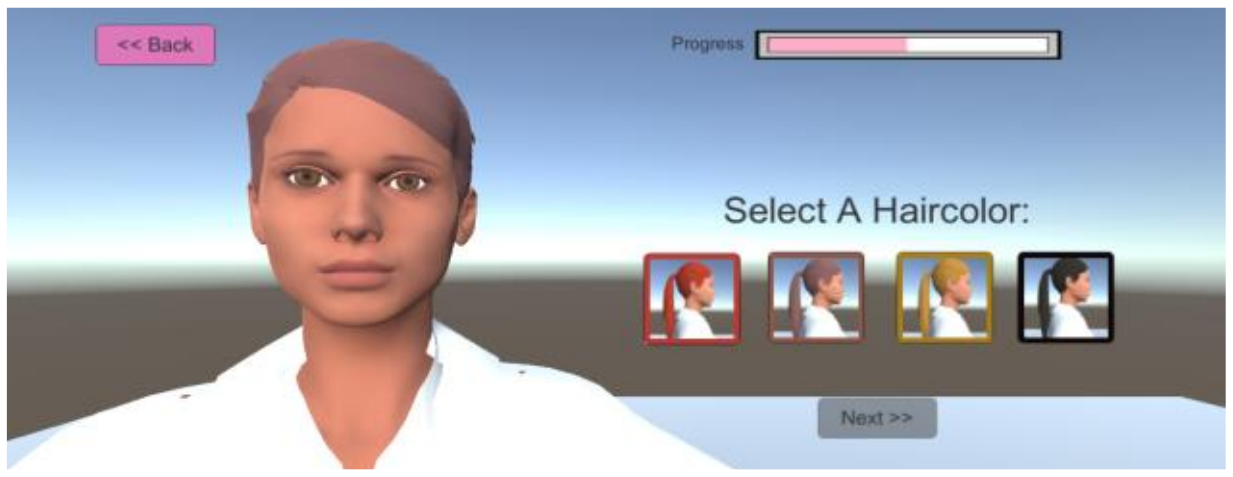

**Fig. 1.** Example customization screen for hair color.

The remainder of this paper is organized as follows. Section II describes the proposed methods. Section III presents the results and discussion. Section IV concludes.

## II. Methods

### *A. Design and Development of CADyFACE Stimuli*

*1) Avatar Generation:* We generate 3D avatars for CADyFACE using free, open-source tools including the 3D modeling software Blender (https://www.blender.org/) and ManuelBastioniLAB 1.6.1a (https://github.com/animate1978/MB-Lab), a character creation plugin for Blender. ManuelBastioniLAB 1.6.1a includes six human prototypes: African female, African male, Asian female, Asian male, European female, and European male. We obtain one avatar for each of these six prototypes using the default settings. Each avatar includes a face rig with 75 blendshapes for facial animation. We dress avatars using clothing assets (shirt, jacket, pants) obtained from [44].

*2) Avatar Customization:* We develop the CADyFACE avatar customization application using the free Unity game engine (https://unity.com/). Users are shown their current avatar on the left side of the screen and a selection of customization options on the right. Users navigate between screens of options using ‘next’ or ‘back’ buttons, which also update a progress bar. As users select different customization options, the updates are rendered on the avatar. An example customization screen is shown in Fig. 1. There are 49,152 different possible combinations based on selection of one of each of the following: six different avatar models, three skin tones, four eye colors, four hair colors, eight jacket colors, eight shirt colors, and eight pants colors. All customization options are summarized in Fig. 2.

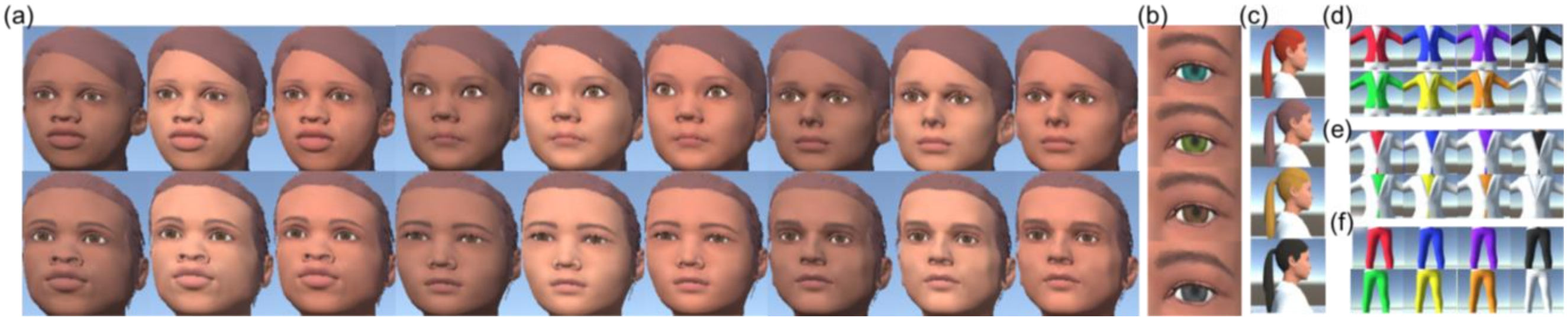

**Fig. 2.** Avatar customization options: (a) all avatar model and skin tone combinations, (b) eye color options, (c) hair color options, (d) jacket color options, (e) shirt color options, and (f) pants color options.

TABLE I

CADyFACE AU Intensities (A=Low to E=High)

| Expressions | 1 | 2 | 4 | 5 | 6 | 7 | 10 | 11 | 12 | 15 | 17 | 20 | 23 | 25 | 26 | 27 |
|---|---|---|---|---|---|---|---|---|---|---|---|---|---|---|---|---|
| **Anger** | | | E | E | | C | D | | | | | | D | B | C | |
| **Disgust** | | | | | | | E | | | | D | | | | | |
| **Fear** | E | C | D | E | | | | | | | | C | | C | | B |
| **Happy** | | | | | C | | | | E | | | | | | | |
| **Sad** | E | | D | | | | | B | | E | | | | | | |
| **Surprise** | D | D | | B | | | | | | | | | | C | | C |

(Column headers: Action Units)

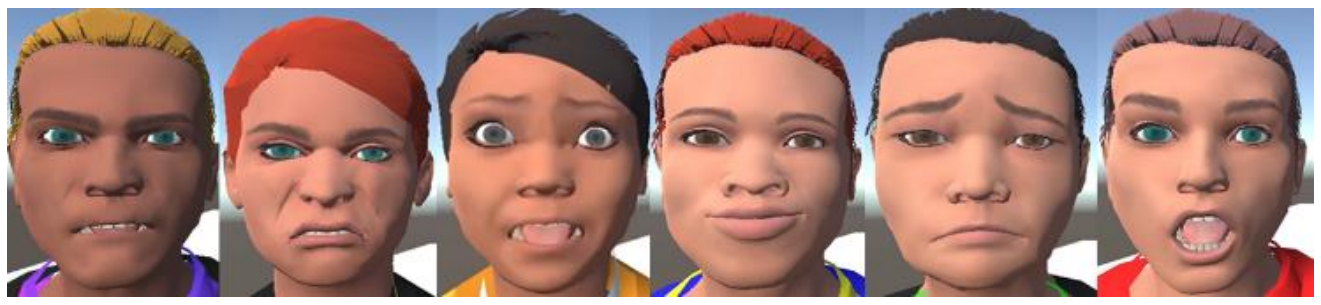

**Fig. 3.** FACS-annotated expressions in CADyFACE (left to right): anger, disgust, fear, happy, sad, and surprise.

*3) FACS-annotated Facial Expressions:* Within Unity, we develop a software application to visualize expressions on each of the six prototype avatars and to adjust the appearance of each expression. Using this software, a member of our team (C. Butler) who is a certified FACS expert with over 600 hours of coding experience has tuned the blendshapes for each of the six prototype avatars to render six different facial expressions (a total of 36 sets of 75 blendshapes). The AUs representing each expression are selected based upon their definitions in the FACS Investigator's Guide [20]. The specific AUs present in each expression and their intensities are reported in Table I. Examples of each expression are shown in Fig. 3.

*4) Dynamic Animation of Facial Expressions:* To generate the facial expression animations for CADyFACE, we linearly interpolate the blendshape values from 0 (neutral) to the values associated with the AU labels defined for the target expression and avatar prototype. We animate each expression over 25 frames with a 50-millisecond delay between frames.

*5) Review by Clinical Team Members:* We have developed CADyFACE as a part of an Institutional Review Board (IRB)-approved study for behavioral biomarker discovery among children and young adults diagnosed with ASD. Two of our team members (J. Keener and J. W. Harrington) who are clinicians with expertise in ASD have reviewed and provided feedback on CADyFACE throughout its development to ensure suitability for the study and appropriateness for individuals diagnosed with ASD.

## *B. BeCoME-Net for Mult-label AU Detection*

*1) Benchmark Data sets:* To train and evaluate BeCoME-Net for detecting the AUs present in CADyFACE, we the consider the Extended Cohn-Kanade (CK+) [59], [60] data set. The CK+ data set comprises 593 image sequences of 123 adult subjects ages 18 to 50 years posing facial expressions including anger, disgust, fear, happy, sad, and surprise. Each sequence begins with a neutral expression frame and ends with the peak expression frame, which has been annotated with AU labels. CK+ includes 30 different AUs, including the 16 AUs in CADyFACE: AUs 1, 2, 4, 5, 6, 7, 10, 11, 12, 15, 17, 20, 23, 25, 26, and 27. We refer to this subset of CK+ AUs as 16AU-CK+. We also use CK+ to compare BeCoME-Net with existing state-of-the-art approaches. However, since some of the AUs in CK+ appear with low frequency, existing state-of-the-art approaches report results on 12 or 13 AU subsets. We follow established literature [28] to define the 12 AU subset (12AU-CK+) as AUs 1, 2, 4, 5, 6, 7, 9, 12, 17, 23, 24, and 25. Then, the 13 AU subset (13AU-CK+) is defined as 12AU-CK+ and AU 27 [28]. The frequencies of AUs present in at least one of these three subsets are reported in Table II. Additionally, we follow the same procedure as [58] to obtain expression-labeled samples of CK+ to train the model to perform the expression classification task as a part of the proposed multi-task learning. The distribution of expression labels is 135 anger, 177 disgust, 75 fear, 207 happy, 327 neutral, 84 sad, and 249 surprise samples.

TABLE II

Frequency of AUs in CK+ and DISFA+ Data Sets

| AU | Description | CK+ Frequency | DISFA+ Frequency |
|---|---|---|---|
| **1** | Inner Brow Raiser | 117 | 9353 |
| **2** | Outer Brow Raiser | 117 | 7982 |
| **4** | Brow Lowerer | 194 | 12036 |
| **5** | Upper Lid Raiser | 102 | 9208 |
| **6** | Cheek Raiser | 123 | 9839 |
| **7** | Lid Tightener | 121 | -- |
| **9** | Nose Wrinkler | 75 | 3993 |
| **10** | Upper Lip Raiser | 21 | -- |
| **11** | Nasolabial Deepener | 34 | -- |
| **12** | Lip Corner Puller | 131 | 10371 |
| **15** | Lip Corner Depressor | 94 | 3956 |
| **17** | Chin Raiser | 202 | 5689 |
| **20** | Lip Stretcher | 79 | 4854 |
| **23** | Lip Tightener | 60 | -- |
| **24** | Lip Pressor | 58 | -- |
| **25** | Lips Part | 324 | 11442 |
| **26** | Jaw Drop | 50 | -- |
| **27** | Mouth Stretch | 81 | 7487 |

In addition to our primary data set, CK+, we also benchmark our approach on the Extended Denver Intensity of Spontaneous Facial Action (DISFA+) [61], [62] data set. DISFA+ consists of image sequences of 9 adult subjects posing 42 facial expressions including individual AUs, combinations of AUs, and 6 basic expressions (anger, disgust, fear, happy, sad, and surprise). Each sequence begins with a neutral expression, moves to the peak expression, and ends with a neutral expression. All frames are annotated for 12 different AUs: 1, 2, 4, 5, 6, 9, 12, 15, 17, 20, 25, and 26. Frequencies of these AUs are reported in Table II. We extract samples with expression labels for use in multi-task learning: 324 anger, 2779 disgust, 1276 fear, 3881 happy, 24793 neutral, 436 sad, and 1345 surprise samples.

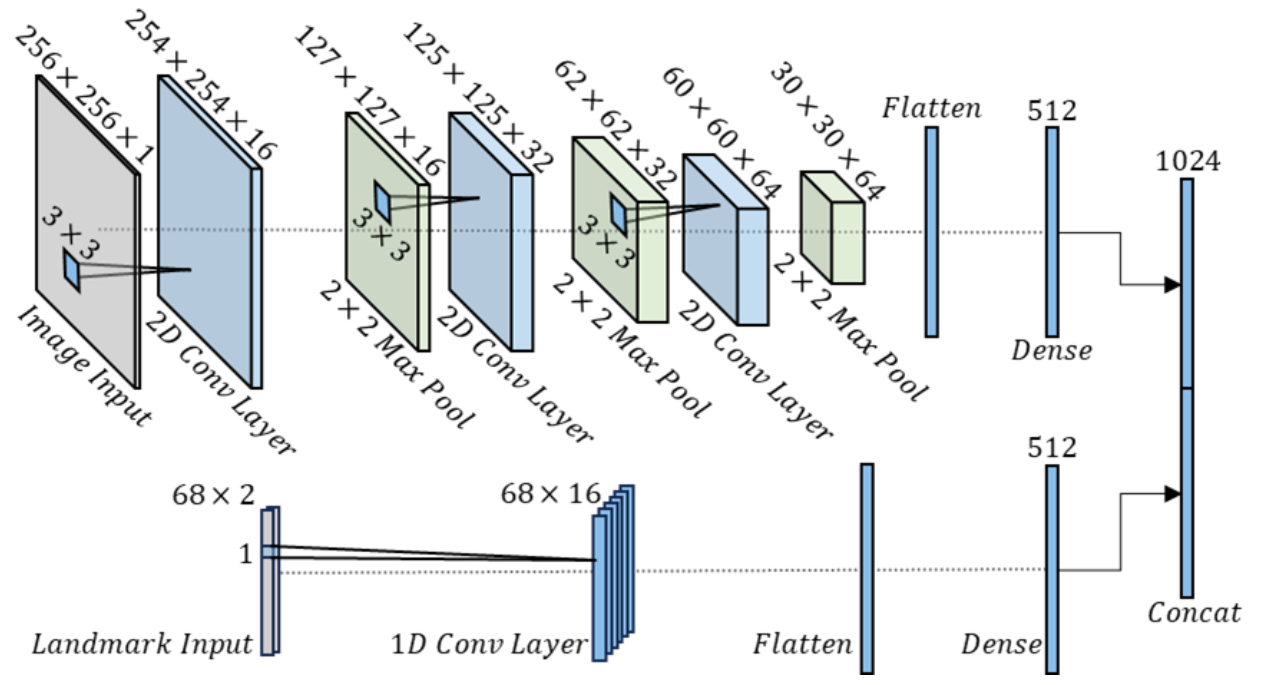

**Fig. 4.** BeCoME-Net-F backbone architecture.

*2) Preprocessing:* We follow the same preprocessing pipeline as in [58], which yields 256×256-pixel grayscale images of centered faces, rotated such that the eyes are level and the left eye is 30% of the image width from the left edge. The images are normalized to the range [0,1]. For each image, we use the dlib library (http://dlib.net/) to extract 68 landmark points on the face and normalize x- and y-coordinates to [0,1].

*3) Notation:* BeCoME-Net is a deep learning model of the form $f: X \rightarrow Y$. We define input $X$ as the tuple $(X_{img}, X_{lmk})$, where $X_{img} \in \mathcal{X}_{img}$ and $X_{lmk} \in \mathcal{X}_{lmk}$. $\mathcal{X}_{img}$ is the set of all $m \times n$ facial expression images, which is a subset of the real number space $\mathbb{R}^{m \times n}$. $\mathcal{X}_{lmk}$ is the set of all pairs of $l$ x- and y-landmark coordinates, which is a subset of $\mathbb{R}^{l \times 2}$. We train BeCoME-Net to perform AU detection and expression classification. For AU detection, we define $Y = Y_{AU} \in \mathcal{Y}_{AU} \subset \mathcal{V}_c$, where $\mathcal{V}_c$ is the set of all $c$-dimensional binary vectors and $\mathcal{Y}_{AU}$ represents the set of all label vectors indicating presence or absence of $c$ different AUs. For expression classification, we define $Y = Y_{EXPR} \in \mathcal{Y}_{EXPR} \subset \mathcal{U}_k$ where $\mathcal{U}_k$ is the set of $k$-dimensional one-hot encoded vectors. We denote the subject identity vector as $g$. We partition $f$ into backbone network $M$ and task head $H$ such that $f = H \circ M$, $M: X \rightarrow Z$, and $H: Z \rightarrow Y$, where $Z$ is a latent space of $p$ features.

*4) Bilateral and Unilateral AU Detection:* We define two variants of BeCoME-Net with different input shapes for bilateral and unilateral detection of AUs. BeCoME-Net-F is designed to process $(m \times n = 256 \times 256)$-pixel grayscale images and $l = 68$ landmark points extracted from the full facial image for bilateral AU detection. For unilateral AU detection, we predict AUs on the left and right sides of the face independently and define BeCoME-Net-H for $(m \times n = 256 \times 128)$-pixel grayscale images of the left or right side of the face and $l = 39$ landmark points (29 from the same side of the face and 10 located along the center line of the face).

*5) Backbone Architecture:* The architecture for BeCoME-Net begins with a backbone $M(\cdot)$, consisting of convolutional, pooling, and fully connected layers for feature extraction. Fig. 4 presents the backbone architecture for BeCoME-Net-F. The backbone incorporates two branches for processing images $X_{img}$ and landmarks $X_{lmk}$, respectively. For the image branch, we consider the same model architecture as in [58]: three blocks of a 2D convolutional layer with 3x3 kernel size followed by 2x2 maximum pooling yielding 16, 32, and 64 feature maps, respectively, and a final fully connected layer of 512 hidden units. Convolutional and fully connected layers use the ReLU activation function. Dropout is applied with a probability of 0.5 at the final fully connected layer. For the landmark branch, we input the x, y-coordinates of the $l$ landmark points directly into a 1D convolutional layer with a kernel size of 1 to yield 16 feature maps, which are flattened prior to a final 512-unit fully connected layer. We use ReLU in convolutional and fully connected layers and apply dropout with a probability of 0.5 at the fully connected layer. Compared to [58] which performs feature engineering and selection based on the landmarks prior to learning, our 1D convolutional layer with kernel size 1 aggregates the 2D coordinate information so that the network may learn relevant features from the normalized landmark positions directly. We concatenate the outputs of the image and landmark branches to form $(p = 1024)$-dimensional feature vector $Z$.

*6) Beta-guided Correlation Loss:* We are interested in modeling significant correlations between the features in $Z$, labels in $Y$, and subject identity $g$ during training. Let $b$ represent the batch size. Consider the space $\mathbb{R}^b$. From [56], [57], [58], the sine squared of the angle $\theta$ between two random lines drawn from $\mathbb{R}^b$ follows the beta distribution:

$$\lambda \stackrel{\text{def}}{=} \sin^2 \theta \sim beta\left(\frac{b-1}{2}, \frac{1}{2}\right) \quad (1)$$

Random or ‘null’ pairs will be approximately perpendicular for even moderate values of $b$, e.g., $b = 10$, meaning that the probability of two random lines being correlated by chance is very small [56]. This result may be used to build a graphical model where the nodes are features, labels, or identity, and edges represent significant correlations. We denote the number of nodes as $w$. For the AU detection task, $w\ nodes = p\ features + c\ AUs + 1\ subject\ identity$. For the expression classification task, $w\ nodes = p\ features + k\ expressions + 1\ subject\ identity$.

To build the graph, we employ a frequentist inferential procedure to screen for edges ('non-null' pairs or significant correlations) among the features in $Z$, labels in $Y$, and subject identity $g$. We denote the $\eta$-quantile of $beta\left(\frac{b-1}{2}, \frac{1}{2}\right)$ as $Q_\eta$. Pairs $\lambda_e$ (e.g., feature-feature, feature-label, feature-identity) are considered significantly correlated if $\lambda_e = \sin^2 \theta_e < Q_\eta$, where $e = 0,1,\dots,t$ and the total number of possible edges $t = 0.5w(w-1)$. The selection of $\eta$ may be used to control the Type I error rate. For each possible edge $\lambda_e$, we consider the null hypothesis $H_0: \lambda_e \geq Q_\eta$ (i.e., no edge) and the alternative $H_a: \lambda_e < Q_\eta$ (i.e., edge in the graph). We conduct a total of $t$ individual hypothesis tests to determine the presence/absence of all possible edges. Using the Bonferroni correction, we divide $\alpha = 0.05$ by the total number of hypothesis tests $t$ to set $\eta = \frac{\alpha}{t}$. The screening rules associated with the null and alternative hypotheses may be implemented using mirrored and translated Heaviside functions:

$$Heaviside(Q_\eta - \lambda_e) = \begin{cases} 0, & \lambda_e \geq Q_\eta \\ 1, & \lambda_e < Q_\eta \end{cases}. \qquad (2)$$

However, due to the discontinuity at $\lambda_e = Q_\eta$, (2) is not differentiable. Sigmoid functions may be used to provide a smooth approximation for the Heaviside functions [63]. Therefore, we consider the following sigmoid function for differentiable implementation of the screening rules:

$$\sigma(\lambda_e) = 1 - \frac{1}{1 + e^{-m(\lambda_e - Q_\eta)}}, \qquad (3)$$

where $m$ adjusts the sharpness of the 1 to 0 transition at $Q_\eta$.

To construct the predicted graph adjacency matrix $A$, we apply (3) for each $\lambda_e$ to yield the edge connection between each $e^{\text{th}}$ pair of nodes (features, labels, or identity) and assign these to the upper triangle of $A$ in row-major order. We fill the diagonal (representing self-connection) with 1's. The lower triangle of $A$ is the upper triangle mirrored over the diagonal. We construct $A$ such that the first $p$ rows and columns represent the $p$ features. The next $c$ or $k$ rows and columns represent the AU or expression labels, respectively. The last row and column represent identity. Then, the beta-guided correlation loss $\mathcal{L}_{BGC}$ is defined as:

$$\mathcal{L}_{BGC}(A) = \frac{1}{w^2} \sum_{i}^{w} \sum_{j}^{w} (S_{ij} \cdot A_{ij}), \qquad (4)$$

where $S$ is a $w \times w$ sign matrix (consisting of -1's, 0's, and 1's) that we use to encourage features to be correlated with the labels, discourage feature correlations with subject identity, and encourage feature diversity by discouraging correlations among the features themselves. We set the diagonal of $S$ to 0's as self-connection will be unchanging and have no impact on the loss. Similarly, labels and identity will not be updated during learning. Only the features will be affected by the gradient updates. Therefore, we multiply the entries of $A$ associated with label-label and label-identity pairs by 0's in $S$ so that they do not contribute to the loss. Since we minimize the loss during learning, rows and columns representing edges between the labels and features are multiplied by -1 to maximize feature correlations with the labels. The remaining entries of $S$ are filled with 1's to discourage correlations with subject identity and among the features. The entries are multiplied by the corresponding entries of $A$. To aggregate the individual loss contributions into a single number, we sum over all entries. Then, we divide by the total number of entries $w^2$ so that the scale of the loss does not change for different numbers of features or labels.

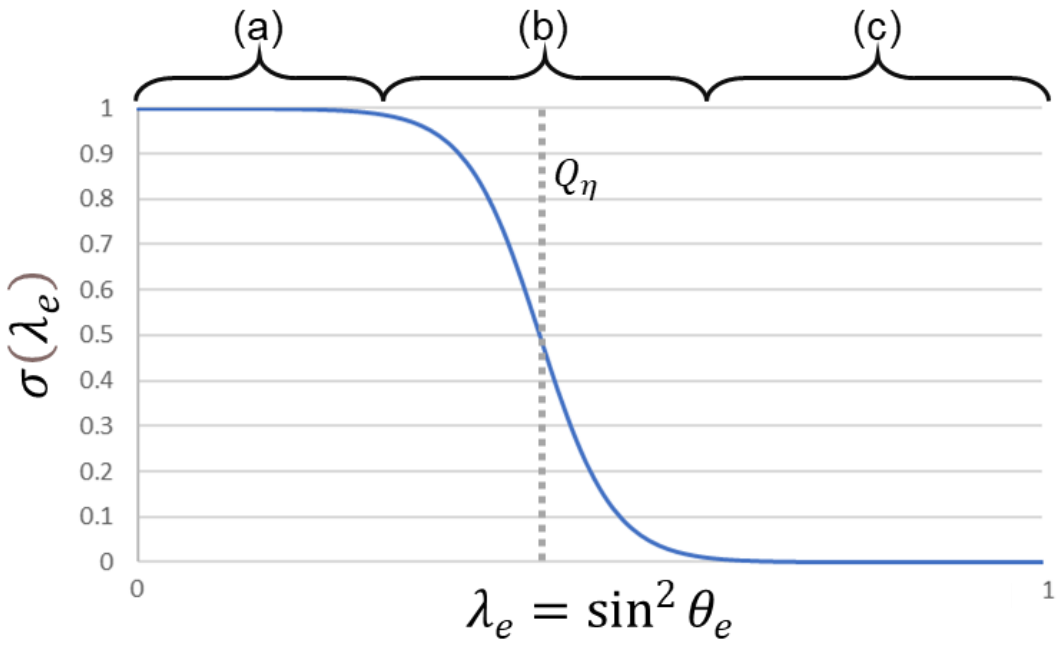

**Fig. 5.** Plot of (3) showing regions where (a) there is a significant correlation between nodes, (b) transition between significantly correlated and uncorrelated nodes, and (c) nodes are uncorrelated.

Equation (4) bears some similarity to reinforcement learning. Considering the policy gradient theorem without discounting [64], the loss at each time step is defined as the immediate reward times the predicted action (e.g., the one-hot encoding for the current action times the log of predicted probabilities for each action). Analogously, our $A_{ij}$'s encode the predicted presence or absence of an edge in the graph and the $S_{ij}$'s encode the associated rewards. However, rather than simply using cross entropy for edge predictions, our (3) has several advantages. Fig. 5 shows three key regions of (3). A particular $\lambda_e$ will fall within region (a) if the associated pair of nodes is significantly correlated and will receive the full reward (or penalty) based on the associated $S_{ij}$. For example, a $\lambda_e$ representing a feature-identity pair with $\sigma(\lambda_e) \approx 1$ will receive a penalty $\approx 1$, while a $\lambda_e$ representing a feature-label pair with $\sigma(\lambda_e) \approx 1$ will receive a reward $\approx 1$ (penalty $\approx -1$). Region (b) represents the boundary between significantly correlated and uncorrelated nodes. For $\lambda_e$'s falling within region (b), the reward will be weighted by $\sigma(\lambda_e)$. Finally, region (c) represents uncorrelated nodes where $\sigma(\lambda_e) \approx 0$, so uncorrelated nodes will have a very small contribution to the loss. Due to (3), $\mathcal{L}_{BGC}$ focuses more on significantly correlated pairs while ignoring uncorrelated pairs. Therefore, individual features are allowed to specialize, i.e., to be highly correlated with one or several specific AUs (or expressions), without being penalized for having low correlation with other AUs (or expressions). This property is especially suitable for AU and expression learning as many of the constituent muscle actions of the face cannot or rarely occur concurrently (e.g., AU 24 'lip pressor' and AU 25 'lips part' cannot occur together) while others often occur together (e.g., AU 1 'inner brow raiser' and AU 2 'outer brow raiser').

*7) Multi-task Learning Framework:* While our primary goal is AU detection, training BeCoME-Net to perform the related task of expression classification is expected to improve representation learning and AU detection performance. We define the multi-label AU detection head as a fully connected output layer of $c$ units with sigmoid activation, where $c$ is the number of target AUs. We define the expression classification head as a fully connected softmax output layer with $k$ units for the $k$ facial expression classes. For efficient learning of both tasks, we duplicate the backbone and connect one head to each copy, as shown in Fig. 6. Weights are shared between the two copies of the backbone for simultaneous training on both AU detection and expression classification tasks. We supervise the learning of the AU detection and expression classification tasks with the weighted multi-label cross-entropy loss $\mathcal{L}_{WMCE}$ [65] and weighted categorical cross entropy loss $\mathcal{L}_{WCCE}$ [66], respectively. We choose weighted variants of both losses to address imbalance in the label distributions. We use the beta-guided correlation loss $\mathcal{L}_{BGC}$ in both tasks ($\mathcal{L}_{BGC_AU}$ for AU detection and $\mathcal{L}_{BGC_EXPR}$ for expression classification) to encourage features to be correlated with the labels while discouraging correlation with subject identity. The loss for the AU detection task is $\mathcal{L}_{AU} = \mathcal{L}_{WMCE} + \mathcal{L}_{BGC_AU}$, and $\mathcal{L}_{EXPR} = \mathcal{L}_{WCCE} + \mathcal{L}_{BGC_EXPR}$ for the expression classification task. The overall loss is $\mathcal{L} = \mathcal{L}_{AU} + \mathcal{L}_{EXPR}$.

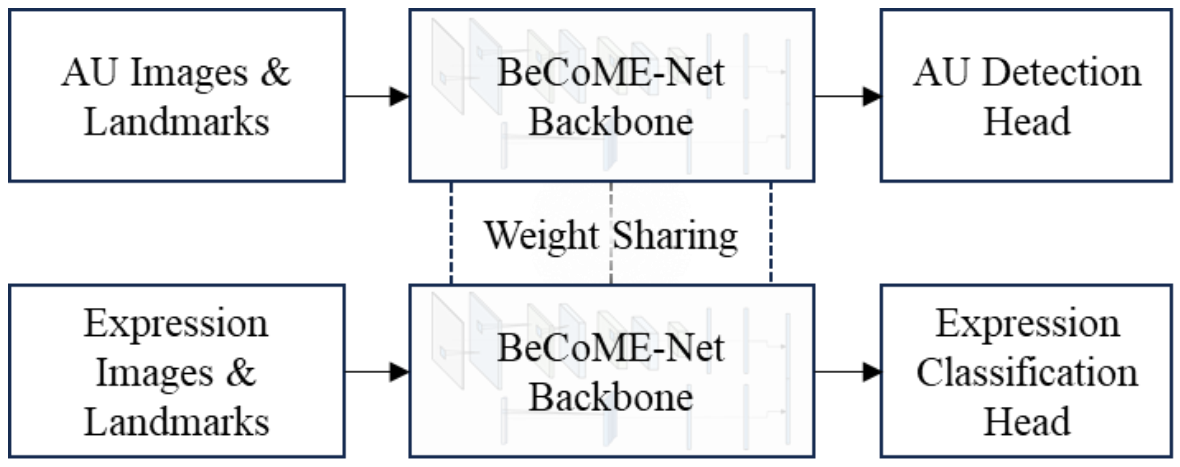

**Fig. 6.** BeCoME-Net multi-task learning framework.

*8) Experiments:* Following established literature [28], we perform 3-fold subject-independent cross validation for all experiments and report F1 scores calculated over all test folds. To study the effect of multi-task learning and the proposed Beta-guided correlation loss, we perform an ablation study on 16AU-CK+ for both bilateral and unilateral AU detection models. Then, we benchmark BeCoME-Net-F and BeCoME-Net-H on 16AU-CK+ and report the performance for each AU. Next, we train and test BeCoME-Net-F and BeCoME-Net-H on 13AU-CK+, 12AU-CK+, and DISFA+ for comparison with state-of-the-art approaches. For 13AU-CK+, we compare with BGCS [25], HRBM [23], and LNDSM [28]. For 12AU-CK+, we compare with JPML [24], DSCMR [22], and LNDSM [28]. For DISFA+, we compare with DRML [21], AU R-CNN [26], JÂA-Net [27], and LNDSM [28]. LNDSM [28] is the newest state-of-the-art approach with which we compare.

For all experiments, we train using the ADAM optimizer with a triangular learning rate policy [67] cycling the learning rate between $10^{-5}$ and $10^{-3}$ until convergence. We use a batch size of $b = 32$ and set $m$ =100.

### C. Feasibility Study, Tasks, and Constructs

*1) Participants:* We have recruited 20 healthy adult volunteers (ages 21 to 35, 4 female) for an online feasibility study of CADyFACE. This feasibility study has been conducted as a part of a larger IRB-approved study to discover behavioral biomarkers for children and young adults diagnosed with ASD and has been approved by the IRBs at Old Dominion University (Application No. 1424272) and Eastern Virginia Medical School (Application No. 19-06-EX-0152). All participants have provided informed consent and have not received compensation for their participation. Inclusion criteria includes being at least 20 years of age at the time of enrollment and having access to an Internet-connected personal computer with a webcam. For privacy, each participant has been assigned a unique subject identifier.

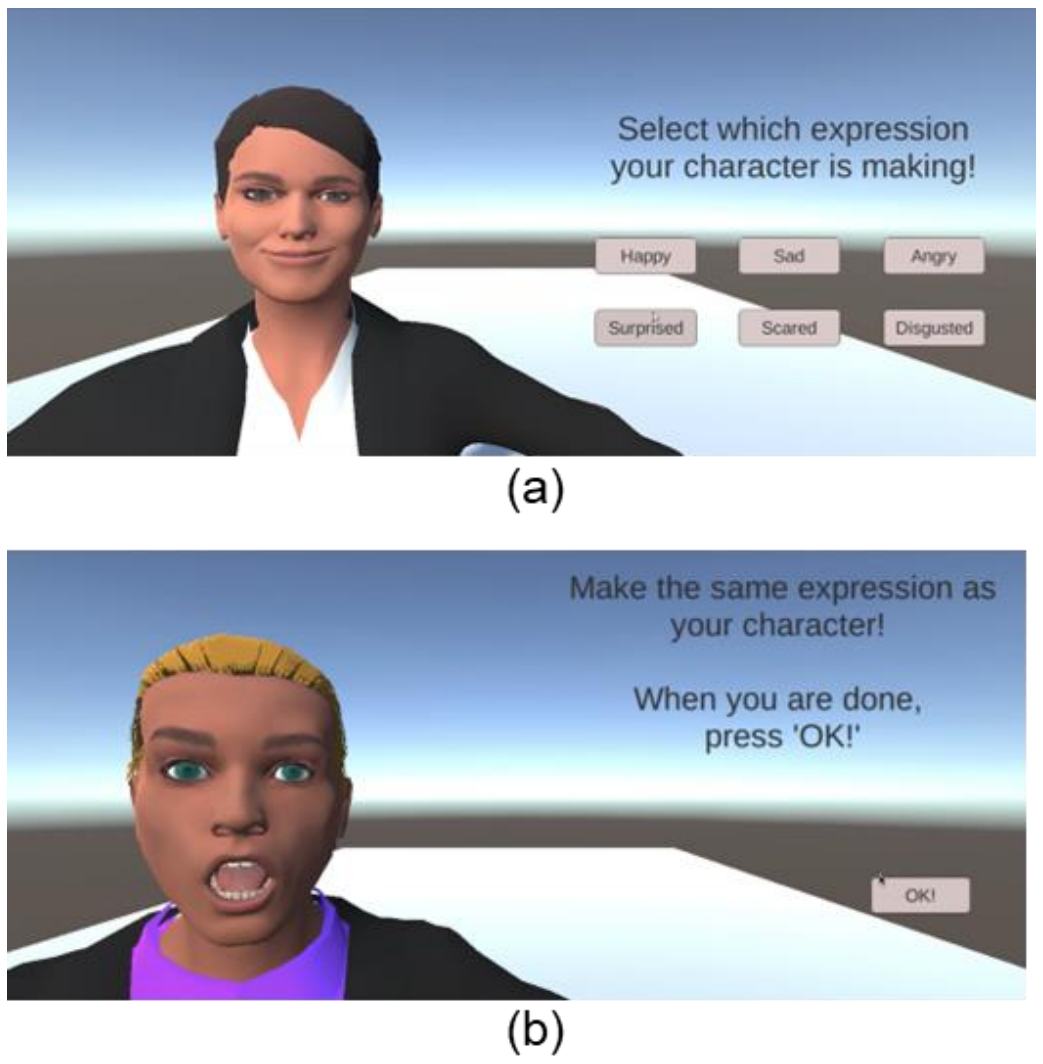

**Fig. 7.** (a) Recognition and (b) mimicry tasks.

*2) Online Stimuli Presentation and Data Collection:* A Unity Web-GL application has been developed to present the CADyFACE stimuli in each participant's web browser. The application is embedded into a webpage hosted on the visionlab.odu.edu domain, which is served by a secure web server located at Old Dominion University. Participants must enter their unique subject identifier to access the webpage.

As the participants interact with the Unity Web-GL application, the WebGazer.js (https://webgazer.cs.brown.edu/) [68] JavaScript library and its self-calibrating eye-tracking model are used to collect video frames and webcam-based eye-tracking coordinates from the participants' webcams and record them to the secure web server.

*3) Tasks:* Participants complete two tasks developed using the CADyFACE stimuli: recognition and mimicry. In the recognition task, participants are asked to select the expression shown by clicking the button labeled with the name of the expression. In the mimicry task, participants are asked to make the same facial expression as the avatar. Each task consists of six trials, one for each of the six FACS-annotated expressions in CADyFACE (anger, disgust, fear, happy, sad, or surprise). Each participant customizes their own avatar for use in both tasks. The expression and mimicry tasks are shown in Fig. 7.

*4) Constructs:* During the recognition task, participants are expected to attend their gaze to the avatar's face to determine the expression. Therefore, we consider the face preference construct [19]. Following [19], we measure the construct as the percentage of gaze duration to the avatar's face (%Face) and test construct validity using a one-sample t-test of %Face against the percentage of the scene taken up by the avatar's face, which is the expected %Face given random gaze. For our recognition task, the avatar's face occupies 15.0% of the scene. We test the construct validity of all six expressions.

During the mimicry task, participants are expected to pose the same facial expression as the avatar. Since CADyFACE has AU labels, we consider constructs based upon the activation of the same AUs by the participants. To measure the constructs, we use BeCoME-Net-F and BeCoME-Net-H to detect the AUs in peak expression frames of each of the participants' mimicked expressions. We use both BeCoME-Net-F and BeCoME-Net-H to measure the construct with both bilateral and unilateral AU detectors. For each AU, we test construct validity using a one-sample t-test against 0 (no activation). We test the construct validity of all AUs in all six expressions.

## III. Results and Discussion

### A. Ablation Study

To understand the impact of multi-task learning and the beta-guided correlation loss, we perform an ablation study for both bilateral and unilateral AU detection on 16AU-CK+. As shown in Table III, the best performance is achieved for bilateral and unilateral models when both multi-task learning and the beta-guided correlation loss are considered. The inclusion of multi-task learning or the beta-guided correlation loss alone result in small improvements in mean F1 score (less than 1%) for bilateral and unilateral models. Including both multi-task learning and beta-guided correlation achieves an improvement of 1.81% and 2.86% in mean F1 score for bilateral and unilateral models, respectively. These results suggest that the use of the beta-guided correlation loss in the secondary expression classification task yields better representation learning for AU detection. These results also show that all bilateral models perform better than their unilateral counterparts, which we expect is due to the bilateral models having access to information from the entire face.

### B. BeCoME-Net Performance for CADyFACE AUs

The precision, recall, and F1 scores for each AU based on 3-fold cross validation of 16AU-CK+ for BeCoME-Net-F and BeCoME-Net-H are reported in Fig. 8. Both models follow similar patterns of performance. The best performing AUs with the F1 scores of over 80% are AUs 2, 12, 17, 25, and 27. As shown in Table II, AUs 2, 12, 17, and 25 are some of the most frequent AUs in 16AU-CK+. While being a less frequent AU, AU 27 (mouth stretch) is associated with the distinctive open mouth appearance seen in the fear and surprise expressions. The worst performing AUs with F1 scores less than 50% are the four least frequent AUs in 16AU-CK+: AUs 10, 11, 23, and 26.

### C. Comparison with State-of-the-art AU Detectors

We compare our proposed BeCoME-Net with state-of-the-art approaches for multi-label AU detection using the CK+ (AU13-CK+ and AU12-CK+) and DISFA+ data sets. Table IV shows that BeCoME-Net-F achieves performance on par with the best performing and most recent state-of-the-art method LNDSM [28] for AU13-CK+. BeCoME-Net-H performs second best after BeCoME-Net-F and LNDSM. For AU12-CK+, BeCoME-Net-F achieves the highest performance, slightly outperforming LNDSM while BeCoME-Net-H performs equally well to LNDSM. We note that for each prediction, LNDSM requires a reference image of the neutral face for the same subject. Our proposed BeCoME-Net performs competitively on CK+ without requiring reference images of the neutral face.

TABLE III
Ablation Study

| AU Detection | Input Size (image, landmarks) | Multi-Task Learning | Beta-Guided Correlation Loss | Mean F1 Score |
|---|---|---|---|---|
| Unilateral | 256x128, 39 | | | 61.16% |
| Unilateral | 256x128, 39 | X | | 61.83% |
| Unilateral | 256x128, 39 | | X | 61.77% |
| ***Unilateral*** | 256x128, 39 | ***X*** | ***X*** | ***64.02%*** |
| Bilateral | 256x256, 68 | | | 64.51% |
| Bilateral | 256x256, 68 | X | | 64.78% |
| Bilateral | 256x256, 68 | | X | 65.16% |
| ***Bilateral*** | 256x256, 68 | ***X*** | ***X*** | ***66.32%*** |

To show how BeCoME-Net performs on a data set other than CK+, we compare our performance on the DISFA+ data set in Table V. While BeCoME-Net-F and BeCoME-Net-H both outperform DRML [21] and AU R-CNN [26], the best performance is achieved by LNDSM [28] followed by JÂA-Net [27]. Both LNDSM and JÂA-Net methods exhibit greater complexity and depth than ours. LNDSM is twice as deep as BeCoME-Net with six convolutional blocks compared to our three [28]. LNDSM also benefits from using neutral reference images to generate saliency maps that are fused at several intermediate layers of the network [28]. JÂA-Net involves multiple sub-networks for face alignment, global feature learning, local AU feature learning, and attention refinement [27]. Furthermore, BeCoME-Net may be less competitive on DISFA+ due to the beta-guided correlation loss, which discourages correlation between the learned features and subject identities. Given that DISFA+ contains only 9 subjects (compared to 123 in CK+), some discriminative features may be spuriously correlated with subject identity.

### D. Construct Validity

As shown in Table VI, the recognition task's face preference construct is valid for all expressions. Two participants are excluded due to tracking loss. Table VII reports construct validity for the mimicry task based on BeCoME-Net-F and BeCoME-Net-H AU predictions. For some of the tests (indicated by an asterisk*), we are unable to compute a t-statistic due to there being no predictions of the AU among any of the participants. These AUs are AU 10, AU 11, AU 23, and AU 26, which are the four least frequent AUs in the 16AU-CK+ training set. The following unilateral and bilateral constructs are valid: AUs 4 and 25 for anger; AU 17 for disgust; AUs 1, 2, 5, 25, and 27 for fear; AU 12 for happy; all AUs (1, 4, 11, 15) except AU 11 for sad; and all AUs (1, 2, 5, 25, 27) for surprise. Failing to pass the test of construct validity for some AUs may be attributed to one of two reasons: BeCoME-Net did not detect a present AU or the CADyFACE stimuli did not successfully elicit the AU.

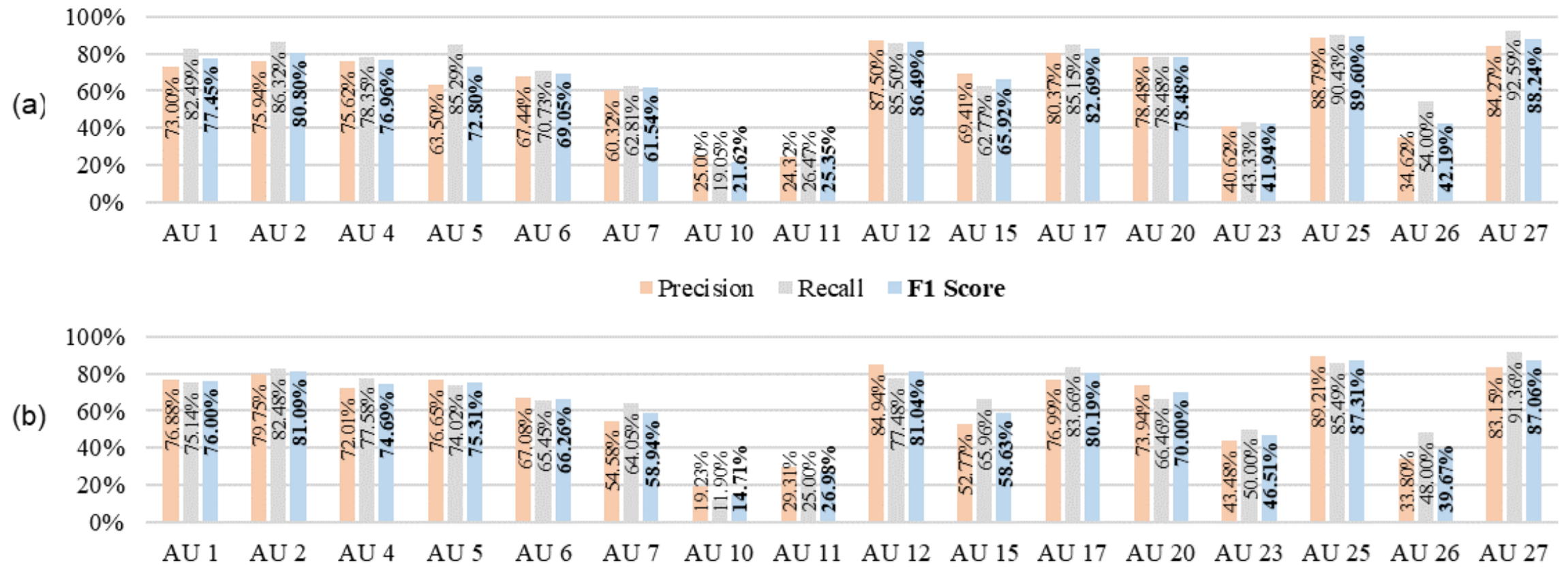

**Fig. 8.** Precision, recall, and F1 scores based on 3-fold cross validation of 16AU-CK+ for (a) BeCoME-Net-F and (b) BeCoME-Net-H.

TABLE IV

F1 SCORES FOR BECOME-NET COMPARED WITH STATE-OF-THE-ART FOR MULTI-LABEL AU DETECTION ON CK+

| Method | AU 1 | 2 | 4 | 5 | 6 | 7 | 9 | 12 | 17 | 23 | 24 | 25 | 27 | Mean |
|---|---|---|---|---|---|---|---|---|---|---|---|---|---|---|
| ***13 AU Detection*** | | | | | | | | | | | | | | |
| BGCS [25] | 0.71 | 0.68 | 0.64 | 0.60 | 0.58 | 0.47 | 0.52 | 0.73 | 0.77 | 0.22 | 0.24 | 0.79 | 0.58 | 0.58 |
| HRBM [23] | **[0.87]** | **0.86** | 0.73 | 0.72 | 0.62 | 0.55 | 0.86 | 0.73 | 0.82 | **[0.57]** | 0.35 | **0.93** | 0.88 | 0.73 |
| LNDSM [28] | **0.86** | **[0.88]** | **[0.81]** | 0.74 | **[0.70]** | **0.62** | 0.89 | **[0.87]** | **[0.86]** | 0.46 | 0.46 | **[0.94]** | **0.90** | **[0.77]** |
| *BeCoME-Net-F* | *0.82* | *0.85* | ***0.80*** | ***[0.76]*** | ***0.66*** | ***[0.63]*** | ***[0.93]*** | *0.82* | ***0.83*** | *0.45* | ***[0.58]*** | *0.91* | ***[0.91]*** | ***[0.77]*** |
| *BeCoME-Net-H* | *0.81* | *0.82* | *0.76* | ***0.75*** | ***0.66*** | *0.58* | ***0.90*** | ***0.83*** | ***0.83*** | ***0.49*** | ***[0.58]*** | *0.90* | ***0.90*** | ***0.76*** |
| ***12 AU Detection*** | | | | | | | | | | | | | | |
| JPML [24] | 0.50 | 0.40 | 0.72 | 0.53 | 0.58 | 0.24 | 0.55 | 0.75 | 0.82 | 0.42 | 0.31 | 0.76 | -- | 0.55 |
| DSCMR [22] | 0.54 | 0.64 | 0.61 | 0.42 | **[0.68]** | 0.36 | 0.54 | 0.80 | **[0.90]** | **[0.75]** | 0.36 | 0.86 | -- | 0.62 |
| LNDSM [28] | **[0.88]** | **[0.86]** | **[0.82]** | 0.75 | **[0.68]** | 0.56 | 0.90 | **[0.87]** | **0.85** | 0.33 | 0.43 | **[0.91]** | -- | **0.74** |
| *BeCoME-Net-F* | *0.80* | ***0.83*** | ***0.80*** | ***0.76*** | ***0.67*** | ***[0.61]*** | ***[0.94]*** | ***0.85*** | *0.82* | ***0.46*** | ***[0.59]*** | ***0.90*** | -- | ***[0.75]*** |
| *BeCoME-Net-H* | ***0.81*** | ***0.83*** | *0.78* | ***[0.78]*** | *0.65* | ***0.60*** | ***0.91*** | *0.82* | *0.81* | *0.44* | ***0.54*** | ***0.90*** | -- | ***0.74*** |

*Bold with brackets indicates the best score. Bold without brackets indicates the second-best score.

TABLE V

F1 SCORES FOR BECOME-NET COMPARED WITH STATE-OF-THE-ART FOR MULTI-LABEL AU DETECTION ON DISFA+

| Method | AU 1 | 2 | 4 | 5 | 6 | 9 | 12 | 15 | 17 | 20 | 25 | 26 | Mean |
|---|---|---|---|---|---|---|---|---|---|---|---|---|---|
| DRML [21] | 0.27 | 0.22 | 0.51 | 0.36 | 0.56 | 0.32 | 0.39 | 0.23 | 0.27 | 0.16 | 0.57 | 0.42 | 0.36 |
| AU R-CNN [26] | 0.48 | 0.43 | 0.56 | 0.48 | 0.43 | 0.24 | 0.47 | 0.24 | 0.06 | 0.29 | 0.53 | 0.39 | 0.38 |
| JÂA-Net [27] | 0.84 | 0.81 | 0.79 | 0.78 | 0.78 | 0.68 | 0.85 | 0.55 | 0.60 | 0.49 | 0.85 | 0.69 | 0.73 |
| LNDSM [28] | 0.83 | 0.80 | 0.78 | 0.74 | 0.82 | 0.74 | 0.84 | 0.56 | 0.65 | 0.50 | 0.88 | 0.77 | 0.74 |
| *BeCoME-Net-F* | *0.75* | *0.71* | *0.70* | *0.68* | *0.72* | *0.64* | *0.81* | *0.53* | *0.60* | *0.41* | *0.74* | *0.60* | *0.66* |
| *BeCoME-Net-H* | *0.73* | *0.74* | *0.67* | *0.69* | *0.72* | *0.61* | *0.73* | *0.39* | *0.49* | *0.37* | *0.69* | *0.55* | *0.61* |

TABLE VI

CONSTRUCT VALIDITY FOR RECOGNITION TASK

| Expression | Construct | %Face df | t-statistic | p-value | validity |
|---|---|---|---|---|---|
| Anger | Face Preference | 17 | 3.523 | 0.001 | ✓ |
| Disgust | Face Preference | 17 | 2.291 | 0.018 | ✓ |
| Fear | Face Preference | 17 | 3.320 | 0.002 | ✓ |
| Happy | Face Preference | 17 | 3.123 | 0.003 | ✓ |
| Sad | Face Preference | 17 | 3.113 | 0.003 | ✓ |
| Surprise | Face Preference | 17 | 2.566 | 0.010 | ✓ |

## IV. LIMITATIONS

As with other 3D avatar models, the ManuelBastioniLAB models that we use in this work are limited by the fidelity and quality of their blendshapes. While AUs 9 and 10 are both listed as potential core components of a prototypical disgust face in the FACS Investigator's Guide [20], AU 9 ('nose wrinkler') is more common, as reflected by the relative frequencies in the CK+ data set. However, since the ManuelBastioniLAB models are unable to perform nose wrinkling, we opt for AU 10. Using AU 9 instead of AU 10 may have yielded better results for the construct validity of the disgust expression. Furthermore, for AU 11 ('nasolabial deepener'), the ManuelBastioniLAB models are only able to render a low level of activation. More conspicuous representation of AU 11 may have had a positive impact on the construct validity of AU 11 within the sad expression.

## V. CONCLUSION

In this article, we propose the CADyFACE stimuli, customizable 3D avatars with FACS labels, intended for use in behavioral biomarker discovery and intervention studies. Additionally, we propose BeCoME-Net for multi-label detection of AUs elicited by CADyFACE. We conduct an online feasibility study with 20 adult volunteers who complete recognition and mimicry tasks based on CADyFACE while their expressions and eye-gaze are recorded. We report construct validity of these tasks using a well-known eye-tracking measure and the BeCoME-Net AU predictions. In the future, we plan to use CADyFACE and BeCoME-Net in a study aimed at discovering behavioral biomarkers for children and young adults with ASD.

TABLE VII
CONSTRUCT VALIDITY FOR MIMICRY TASK

| Expression | Construct | BeCoME-Net-F (Bilateral) | | | | BeCoME-Net-H (Unilateral) | | | |
|---|---|---|---|---|---|---|---|---|---|
| | | df | t-statistic | p-value | validity | df | t-statistic | p-value | validity |
| Anger | AU 4 Activation | 19 | 3.199 | 0.005 | ✓ | 19 | 4.359 | <0.001 | ✓ |
| | AU 5 Activation | 19 | 1.831 | 0.082 | | 19 | 1.831 | 0.083 | |
| | AU 7 Activation | 19 | 1.453 | 0.163 | | 19 | 3.199 | 0.005 | ✓ |
| | AU 10 Activation | 19 | --* | -- | | 19 | 2.517 | 0.021 | ✓ |
| | AU 23 Activation | 19 | --* | -- | | 19 | --* | -- | |
| | AU 25 Activation | 19 | 5.339 | <0.001 | ✓ | 19 | 3.943 | <0.001 | ✓ |
| | AU 26 Activation | 19 | --* | -- | | 19 | 1.453 | 0.163 | |
| Disgust | AU 10 Activation | 19 | 1.000 | 0.330 | | 19 | 1.453 | 0.163 | |
| | AU 17 Activation | 19 | 4.819 | <0.001 | ✓ | 19 | 2.854 | 0.010 | ✓ |
| Fear | AU 1 Activation | 19 | 7.550 | <0.001 | ✓ | 19 | 8.718 | <0.001 | ✓ |
| | AU 2 Activation | 19 | 13.077 | <0.001 | ✓ | 19 | 8.718 | <0.001 | ✓ |
| | AU 4 Activation | 19 | 1.831 | 0.083 | | 19 | 2.517 | 0.021 | ✓ |
| | AU 5 Activation | 19 | 3.943 | <0.001 | ✓ | 19 | 3.943 | <0.001 | ✓ |
| | AU 20 Activation | 19 | 1.000 | 0.330 | | 19 | 1.831 | 0.083 | |
| | AU 25 Activation | 19 | 10.376 | <0.001 | ✓ | 19 | 8.718 | <0.001 | ✓ |
| | AU 27 Activation | 19 | 3.199 | 0.005 | ✓ | 19 | 2.854 | 0.010 | ✓ |
| Happy | AU 6 Activation | 19 | 2.179 | 0.042 | ✓ | 19 | 1.000 | 0.330 | |
| | AU 12 Activation | 19 | 2.854 | 0.010 | ✓ | 19 | 2.854 | 0.010 | ✓ |
| Sad | AU 1 Activation | 19 | 3.943 | <0.001 | ✓ | 19 | 4.818 | <0.001 | ✓ |
| | AU 4 Activation | 19 | 4.359 | <0.001 | ✓ | 19 | 5.338 | <0.001 | ✓ |
| | AU 11 Activation | 19 | --* | -- | | 19 | --* | -- | |
| | AU 15 Activation | 19 | 5.339 | <0.001 | ✓ | 19 | 2.517 | 0.021 | ✓ |
| Surprise | AU 1 Activation | 19 | 4.359 | <0.001 | ✓ | 19 | 5.339 | <0.001 | ✓ |
| | AU 2 Activation | 19 | 4.819 | <0.001 | ✓ | 19 | 5.940 | <0.001 | ✓ |
| | AU 5 Activation | 19 | 2.854 | 0.010 | ✓ | 19 | 3.943 | <0.001 | ✓ |
| | AU 25 Activation | 19 | 7.550 | <0.001 | ✓ | 19 | 10.376 | <0.001 | ✓ |
| | AU 27 Activation | 19 | 3.560 | 0.002 | ✓ | 19 | 3.199 | 0.005 | ✓ |

## ACKNOWLEDGEMENT

The authors would like to express their gratitude to all volunteers who participated in the feasibility study; to Abigail Stedman for her assistance with avatar generation, avatar customization, and task implementation; to Old Dominion University Information Technology Services for their assistance with the web server used for the feasibility study; and to Gregory Hubbard and Elija Bullock for their assistance with eye-tracking data collection and visualization.

**Megan A. Witherow** received the B.S. degree in computer engineering from Old Dominion University (ODU), Norfolk, VA, USA in 2018. She is currently a PhD candidate at the Vision Laboratory, Dept. of Electrical and Computer Engineering, ODU, and a 2020 NSF Graduate Research Fellow. Her research interests include computer vision, deep learning, human-computer interaction, and affective computing.

**Crystal Butler** received the Ph.D. in computer science from New York University (NYU) in 2021. Her academic research focused on applying natural language processing techniques to crowdsourced label sets, which were provided by human annotators in response to generatively modeled synthetic facial expression imagery. She has over 600 hours of experience analyzing videos and images with the Facial Action Coding System (FACS), a comprehensive and widely used system for describing facial expressions. Currently, she works for Mayo Clinic on the development of generative artificial intelligence applications in the field of radiology.

**Winston J. Shields** received the M.S. degree in computer science from Old Dominion University (ODU), Norfolk, VA, in 2022. He is currently a full stack software engineer at the company CoStar Group.

**Furkan Ilgin** received the BS degree in electrical engineering from the Old Dominion University, Norfolk, VA, USA in 2023 and the MBA degree from Western Governors University, Millcreek, UT. He is currently working toward the ME degree in Systems Engineering from University of Virginia, Charlottesville, VA. His research interests include machine learning, renewable energy, energy storage systems, microgrid design, engineering management, risk analysis and human technology interaction.

**Norou Diawara** is Professor of Statistics in the Mathematics and Statistics Department at Old Dominion University, Norfolk, VA, USA. Prof. Diawara received his B.S. at the University Cheick Anta Diop in Dakar, Senegal; Maîtrise in Mathematics at University of Le Havre, France; Master’s in Statistics at University South Alabama; and Ph.D. in Statistics at Auburn University, AL in 2006. His research areas are in estimation techniques of time to event data analyses and neighborhood level causal effects. Such research interests may be included in choice models, statistical pattern recognition using copulas and spatial-temporal models.

**Janice Keener**, PsyD, has extensive training in the assessment of Autism Spectrum Disorder. She is a Certified Trainer in the administration of the Autism Diagnostic Observation Schedule-Second Edition and obtained her research reliability from the Center for Autism and the Developing Brain. Dr. Keener's research and clinical interests include Pediatric Health Psychology, early childhood assessment, consultation and liaison, and Autism Spectrum Disorder. She is certified in the treatment of Tourette Disorder from the Tourette Syndrome Behavioral Therapy Institute Program. She is also bi-lingual and provides assessment and psychotherapy in Spanish and English.

**John W. Harrington**, MD, FAAP is Professor of Pediatrics and the long-time Division Director of General Academic Pediatrics at Eastern Virginia Medical School and Children’s Hospital of The King’s Daughters (CHKD) in Norfolk, VA, USA. He currently is the new Vice-President of Quality/Safety and Clinical Integration at CHKD. He has received multiple grants and is widely published in his main areas of interest: autism, obesity, and vaccine delivery.

**Khan M. Iftekharuddin** received the B.Sc. degree in electrical and electronic engineering from the Bangladesh Institute of Technology, Dhaka, Bangladesh, in 1989, and the M.S. and Ph.D. degrees in electrical and computer engineering from the University of Dayton, Dayton, OH, USA, in 1991 and 1995 respectively. He is a professor and Batten Endowed Chair in Machine Learning in the department of Electrical and Computer Engineering at Old Dominion University (ODU), Norfolk, VA, USA, and the Director of the ODU Vision Laboratory. His research interests include computational modeling, machine learning, medical image analysis, omics data analysis, distortion-invariant recognition, biologically inspired human and machine centric recognition, and computer vision.